\documentclass[english,a4paper]{revtex4}
\usepackage[T1]{fontenc}
\usepackage[latin1]{inputenc}
\usepackage{float}
\usepackage{amstext}
\usepackage{graphicx}

\makeatletter
\@ifundefined{textcolor}{}
{%
 \definecolor{BLACK}{gray}{0}
 \definecolor{WHITE}{gray}{1}
 \definecolor{RED}{rgb}{1,0,0}
 \definecolor{GREEN}{rgb}{0,1,0}
 \definecolor{BLUE}{rgb}{0,0,1}
 \definecolor{CYAN}{cmyk}{1,0,0,0}
 \definecolor{MAGENTA}{cmyk}{0,1,0,0}
 \definecolor{YELLOW}{cmyk}{0,0,1,0}
 }

\makeatletter

\AtBeginDocument{
  
}

\makeatother
 \usepackage{subfigure}

\makeatother

\usepackage{babel}
\begin{document}

\title{Monte Carlo simulations in a disordered binary Ising model}

\author{D. S. Cambuí}

\email{dcambui@fisica.ufmt.br}

\selectlanguage{english}%

\author{A. S. de Arruda }

\email{aarruda@fisica.ufmt.br}

\selectlanguage{english}%

\author{M. Godoy}

\email{mgodoy@fisica.ufmt.br}

\selectlanguage{english}%

\affiliation{Instituto de Física, Universidade Federal de Mato Grosso, 78060-900,
Cuiabá, Mato Grosso, Brazil.}
\begin{abstract}
In this work we study a disordered binary Ising model on the square
lattice. The model system consists of two different particles with
spin-1/2 and spin-1, which are randomly distributed on the lattice.
It has been considered only spin nearest-neighbor exchange interactions
with $J>0$. This system can represent a disordered magnetic binary
alloy $A_{x}B_{1-x}$, obtained from the high temperature quenching
of a liquid mixture. The results were obtained by the use of Monte
Carlo simulations for several lattice sizes $L$, temperature $T$
and concentration $x$ of ions $A$ with spin-1/2. We found its critical
temperature, through the reduced fourth-order Binder cumulant for
the several values of the concentration $x$ of particles (spin-1/2,
spin-1), and also the magnetization, the susceptibility and the specific
heat as a function of temperature $T$. 
\end{abstract}
\maketitle

\section{Introduction}

Magnetic properties of binary site-substitutionally disordered Ising
models have been receiving considerable attention from both theoretical
and experimental point of view \cite{stinchcombe}. In these systems,
two different types of magnetic ions (denoted by $A$ and $B$) are
randomly distributed on a lattice representing a magnetic binary alloy
$A_{x}B_{1-x}$, which is suddenly frozen from high (liquid state)
to low temperatures (solid state). Moreover, this systems may exhibit
a rich diagram phase.

Much of the theoretical work has assumed the two magnetic ions having
the same spin-1/2 value. These models have been investigated by mean-field
approches \cite{thorpe,katsura,kaneyoshi1} and Monte Carlo simulation
\cite{scholten}. On the other hand, less attention has been given
to binary random-site systems where the constituints have different
spin values. Thefore, it is interesting to investigate a system as
the binary random-site Ising model with one of the constituints having
spin-1 and the other having spin-1/2. These systems were investigated
by mean-field theories \cite{kaneyoshi2,kaneyoshi3,plascak} and Monte
Carlo simulation \cite{godoy}. In the reference \cite{godoy} the
authors studied the system with half of the lattice with spin-1/2
and another half with spin-1 randomly distributed.

In this work, we used the mixed-spin Ising model approach for a binary
alloy. The results were obtained by the use of Monte Carlo simulations
for several lattice sizes $L$, temperature $T$ and for several values
of the concentration $x$ of ions $A$ with spin-1/2. We found its
critical temperature, through the reduced fourth-order Binder cumulant
for the several values of the concentration $x$ of the particles
(spin-1/2, spin-1), and also the magnetization, the susceptibility
and the specific heat as a function of temperature $T$. 

The paper is organized as follows: in Section II, we describe the
disordered binary Ising model and we define some observables of interest.
In Section III, we present some details concerning the simulation
procedures and the results obtained. Finally, in the last Section,
we present our conclusions.

\section{model and some observables of interest}

The configurational energy of disordered binary Ising model may be
described by the following Hamiltonian

\begin{equation}
\mathcal{H}=-J\sum_{\left\langle i,j\right\rangle }\{c_{i}c_{j}\sigma_{i}\sigma_{j}+(1-c_{i})(1-c_{j})S_{i}S_{j}\textrm{ }+c_{i}(1-c_{j})\sigma_{i}S_{j}+c_{j}(1-c_{i})S_{i}\sigma_{j}\}\textrm{,}\label{eq:1}
\end{equation}
 where the spin variables assume the values $S_{i}=\pm1,0$ and $\sigma_{j}=\pm1/2$,
and the nearest-neighbor interaction is ferromagnetic, $J>0$. We
associated to each site $i$ of the lattice a occupation variable
$c_{i}$, so that $c_{i}=1$, if the site is occupied by a particle
with spin $\sigma=1/2$ and $c_{i}=0$, if it is occupied by a particle
with spin $S=1$. The sites are occupied independently by the two
particles, with a probability distribution defined as

\begin{equation}
P(c_{i})=\left[(1-x)\delta_{c_{j},0}+x\delta_{c_{j},1}\right],\;\;\;\; i=1,2,\ldots,N\textrm{ .}\label{eq:2}
\end{equation}
where $x$ is the concentration of spin-1/2 and $(1-x)$ is the concentration
of spin-1.

We calculate the following thermodynamic quantities per site: the
magnetization $m_{L}$
\begin{equation}
m_{L}=\left[\frac{1}{N}\left\langle \left|\sum_{i=1}^{N}\left\{ (1-c_{i})S_{i}+c_{i}\sigma_{i}\right\} \right|\right\rangle \right]\textrm{ ,}\label{eq:3}
\end{equation}
the energy $E_{L}$ 
\begin{equation}
E_{L}=\frac{1}{N}\left[\left\langle \mathcal{H}\right\rangle \right]\textrm{ ,}\label{eq:4}
\end{equation}
the susceptibility $\chi_{L}$ 
\begin{equation}
\chi_{L}=N\left(\left[\left\langle m_{L}^{2}\right\rangle \right]-\left[\left\langle m_{L}\right\rangle ^{2}\right]\right)\textrm{ ,}\label{eq:5}
\end{equation}
and the specific heat $c_{L}$ 
\begin{equation}
c_{L}=N\left(\left[\left\langle E_{L}^{2}\right\rangle \right]-\left[\left\langle E_{L}\right\rangle ^{2}\right]\right)\textrm{ }.\label{eq:6}
\end{equation}
 To find the critical point, we used the fourth-order Binder cumulant
\cite{heermann,binder}, 
\begin{equation}
U_{L}=1-\frac{\left[\left\langle m_{L}^{4}\right\rangle \right]}{3\left[\left\langle m_{L}^{2}\right\rangle ^{2}\right]}\textrm{ }.\label{eq:7}
\end{equation}
 In the expressions above $\left[\cdots\right]$ denotes average over
the samples of the system, and $\left\langle \cdots\right\rangle $
denotes the thermal average. $N=L^{2}$ is the total number of particles.

\section{Procedures and results}

In order to determine the critical behavior of this disordered binary
Ising model we employed Monte Carlo simulation techniques. We consider
a square lattice of linear size $L$, with values of $L$ ranging
from $L=12$ to $L=48$, and we applied periodic boundary conditions.

\begin{figure}[H]
\hfill{}\centering\subfigure[]{\label{fig:First-subfigure}\includegraphics[scale=0.3]{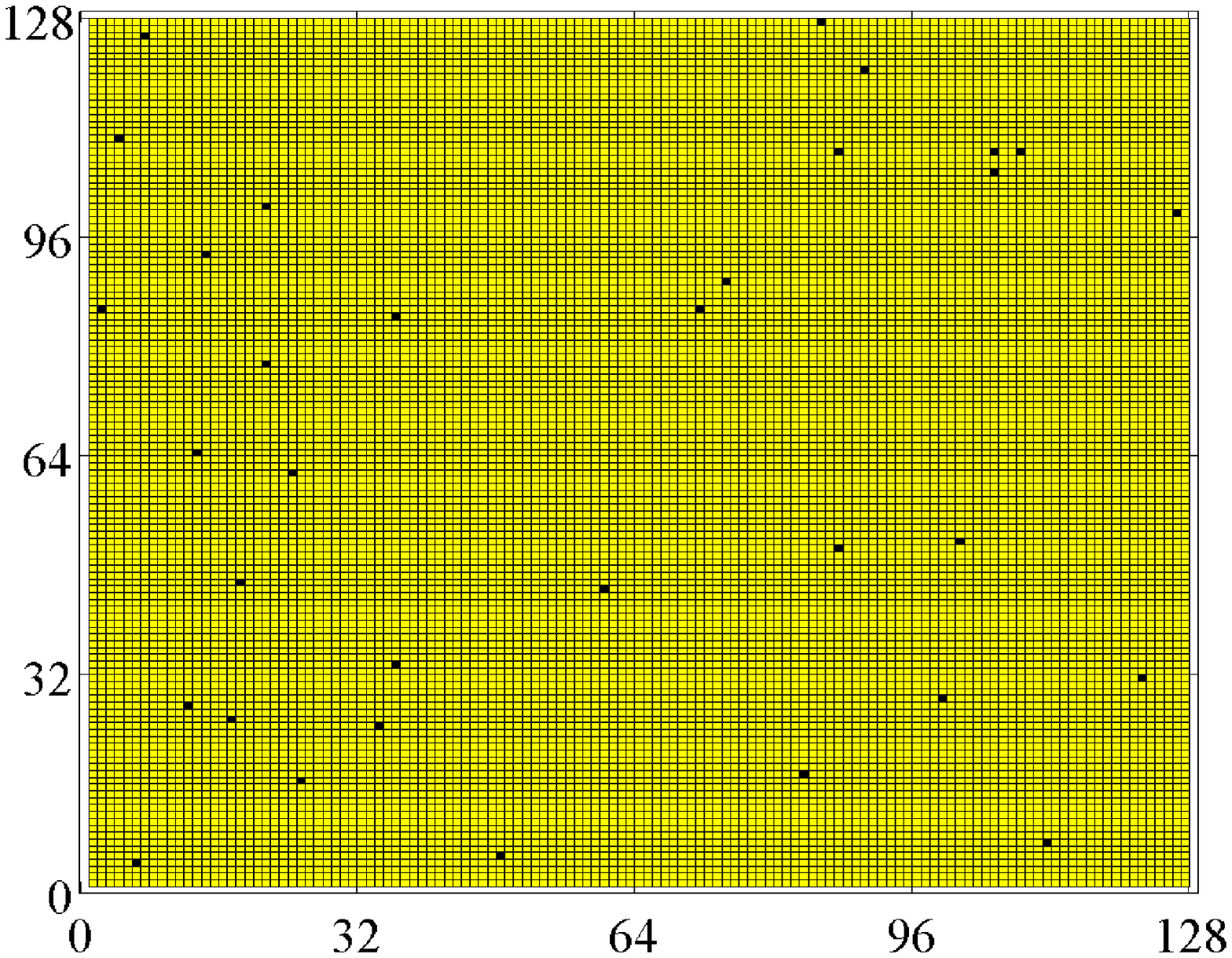}}\hfill{}\subfigure[]{\label{fig:Second-subfigure}\includegraphics[scale=0.3]{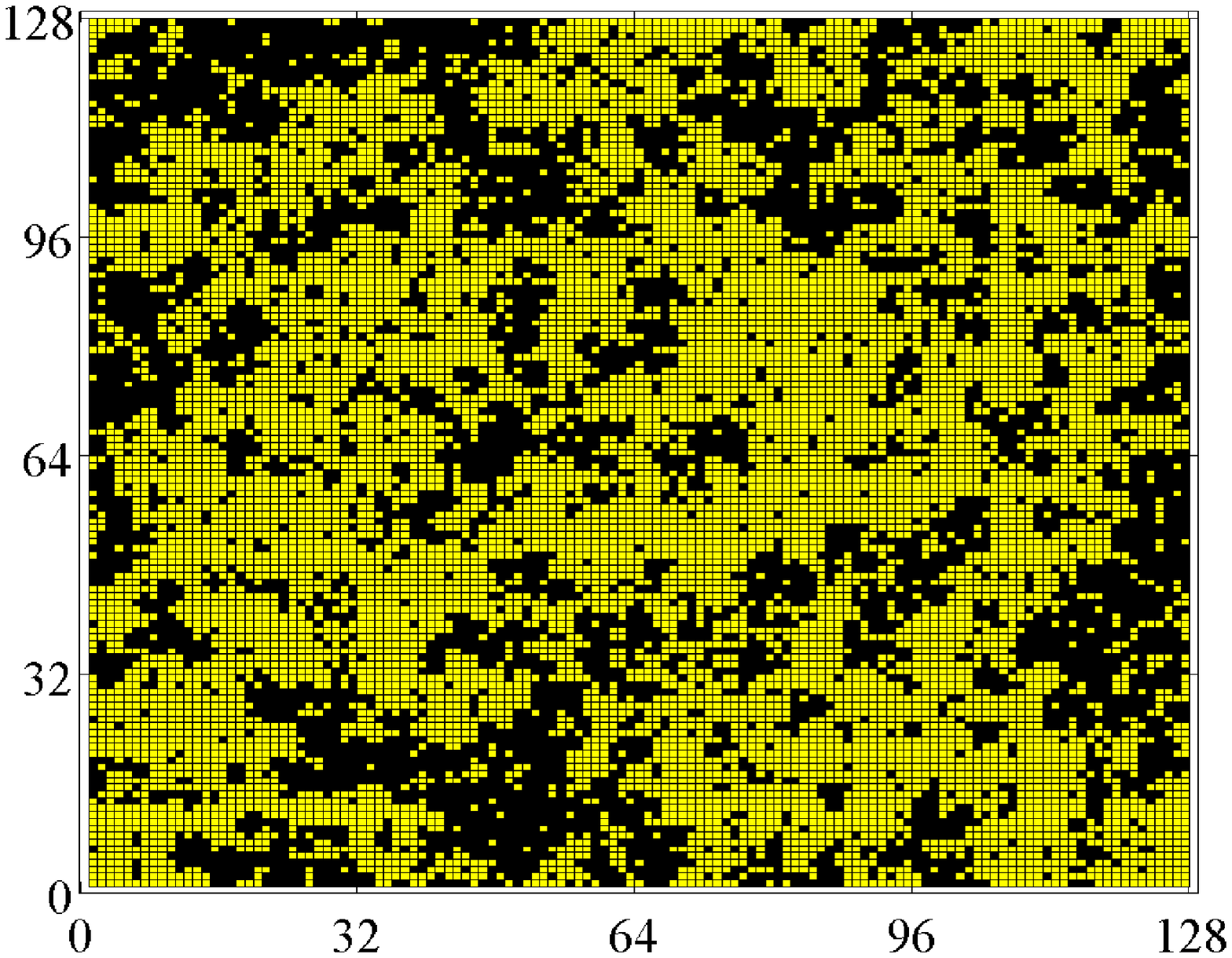}}\hfill{}

\hfill{}\centering\subfigure[]{\label{fig:First-subfigure-1}\includegraphics[scale=0.3]{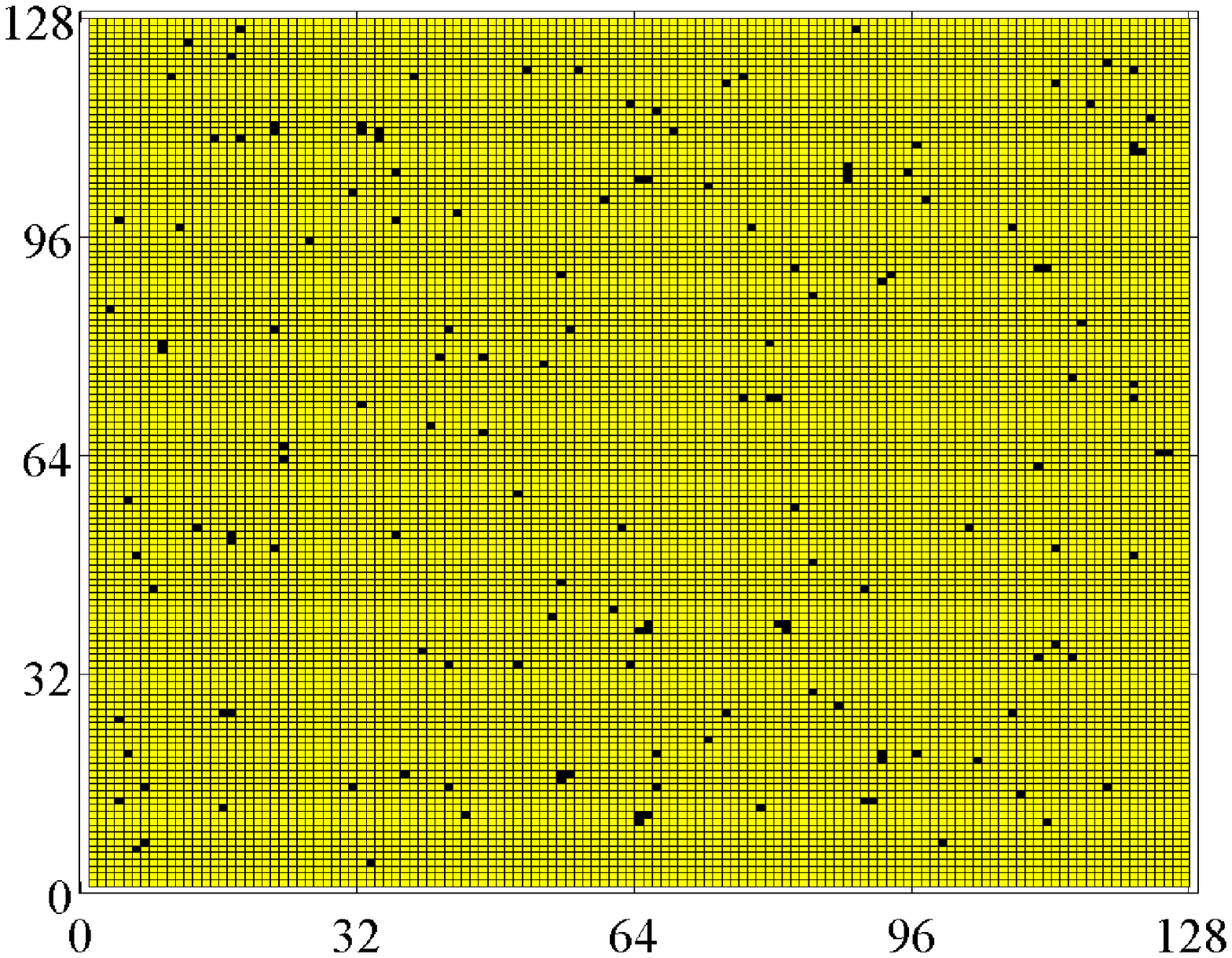}}\hfill{}\subfigure[]{\label{fig:Second-subfigure-1}\includegraphics[scale=0.3]{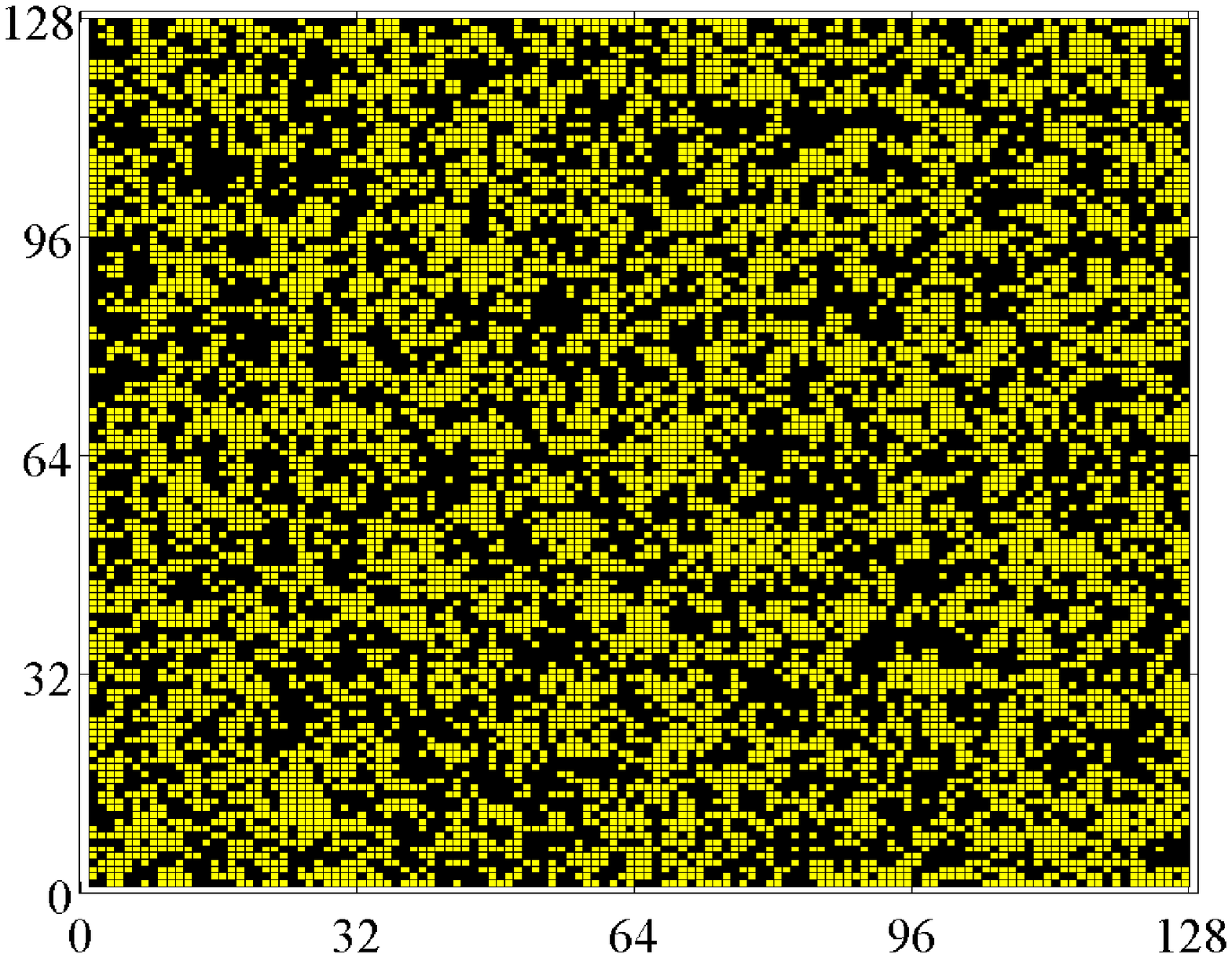}}\hfill{}

\caption{{\small Snapshots of the spin confi{}gurations are given for two different
concentrations $x$ of spin-1/2 and two different values of temperatures
$T$. For $x=0.25$, $T=0.5$ (a) and $T=1.5$ (b); $x=0.75$, $T=0.5$
(c) and $T=1.5$ (d).} spin-1/2 (spin-1) are represented by yellow
(black). {\small The temperature is measured in units of $J/k_{B}$. }}
\end{figure}

We prepared the system with the spins randomly distributed on the
lattice. The concentration $x$ of spin-1/2 and $(1-x)$ spin-1, is
fixed. However, each spin-1/2 can have as its nearest-neighbor spins
of the type spin-1/2 or spin-1 and vice versa. Each trial change of
a spin state on the lattice is accepted according to the Metropolis
prescription \cite{metropolis}, $\omega(\alpha\rightarrow\alpha^{\prime})=\min[1,\exp(-\beta\Delta E)]$,
where $\Delta E$ is the local energy change $(E_{\alpha^{\prime}}-E_{\alpha})$
resulting from changing the state of a random selected spin from $\alpha$
to $\alpha^{\prime}$ state, and $\beta=1/k_{B}T$. To reach the equilibrium
state we take for guarantee at least $1\times10^{6}$ MCs (Monte Carlo
steps) for all the lattice sizes we studied. Then, we take more $5\times10^{5}$
MCs to estimate the average values of the quantities of interest.
Here, 1 MCs means $L^{2}$ trials to change the state of a spin of
the lattice. The average over the disorder was done by using 100 independent
samples for lattices in the range $12\leq L\leq48$.

In Fig. 1, we have a snapshot of the concentrations of spin-1/2 and
spin-1, and they were made for two different concentrations $x=0.25$
and $x=0.75$ of spin-1/2. In both cases, it has been analyzed the
different concentrations at low temperatures ($T<T_{c}$, $T=0.5$)
as shown in Fig. 1(a) and Fig. 1(c), and at high temperatures ($T>T_{c}$,
$T=1.5$) as shown in Fig. 1(b) and Fig. 1(d). {\small The temperature
is measured in units of $J/k_{B}$ throughout the paper. }{\small \par}

\begin{figure}[H]
\hfill{}\centering\subfigure[]{\label{fig:First-subfigure-2}\includegraphics[scale=0.5]{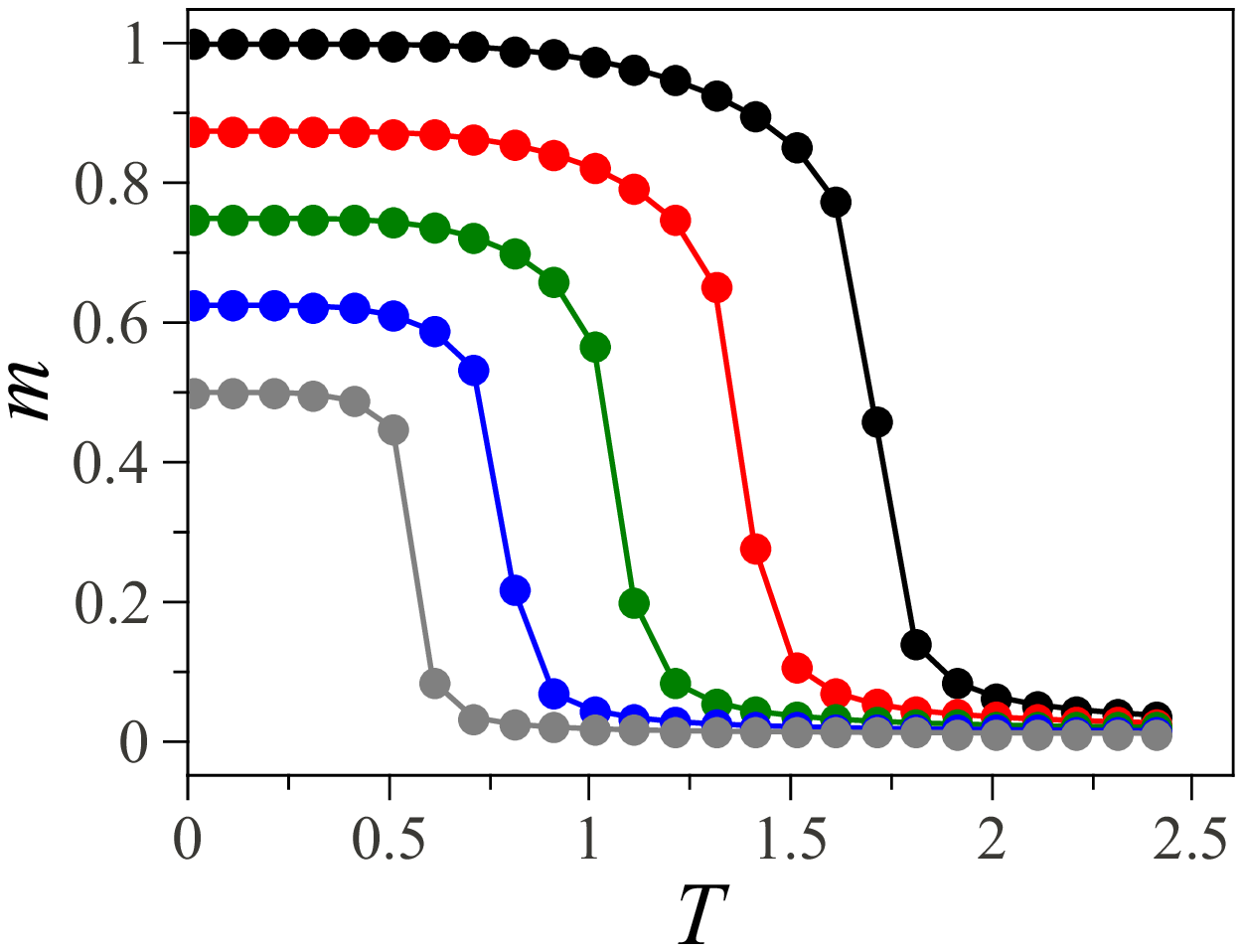}}\hfill{}\subfigure[]{\label{fig:Second-subfigure-2}\includegraphics[scale=0.5]{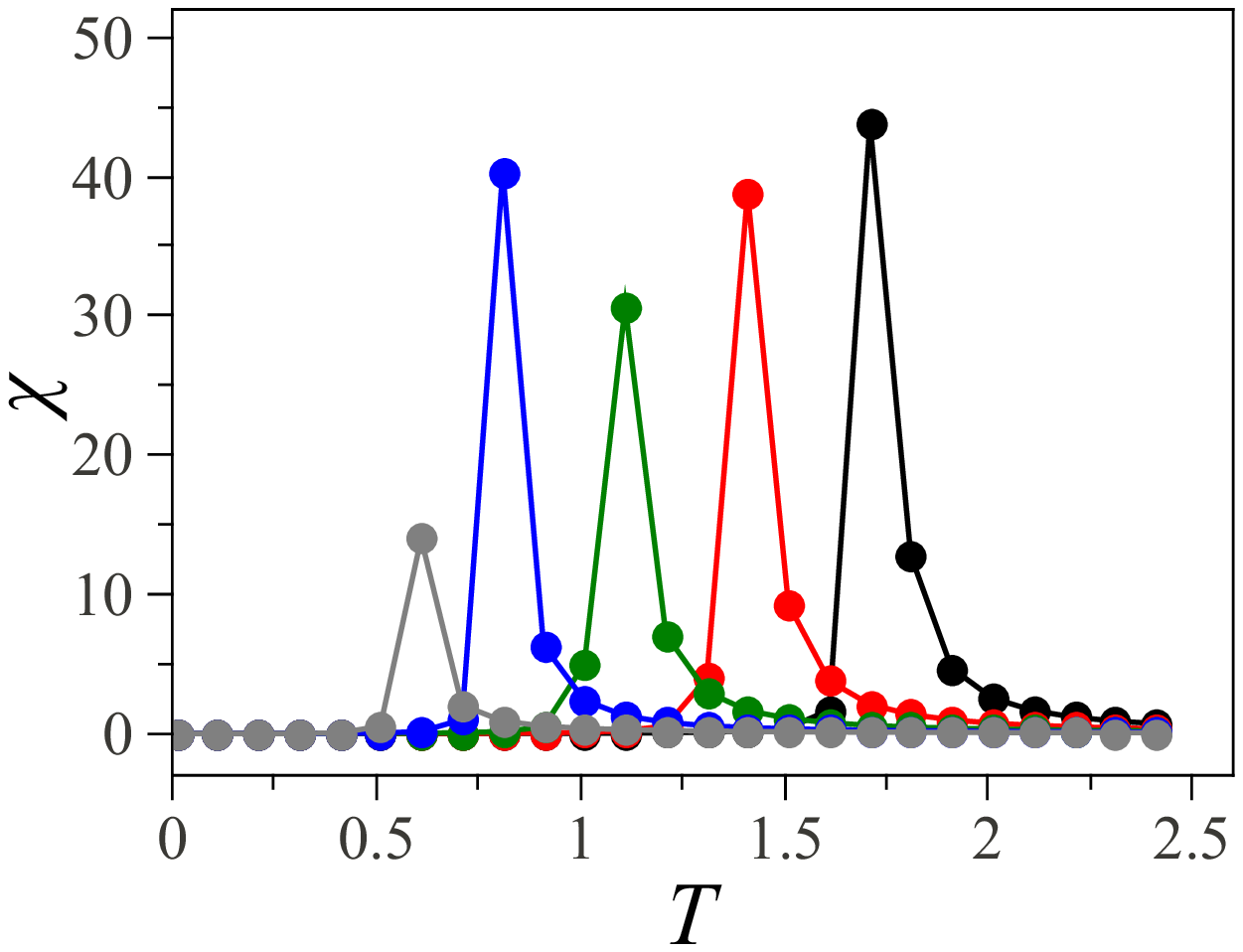}}\hfill{}

\hfill{}\centering\subfigure[]{\label{fig:First-subfigure-1-1}\includegraphics[scale=0.5]{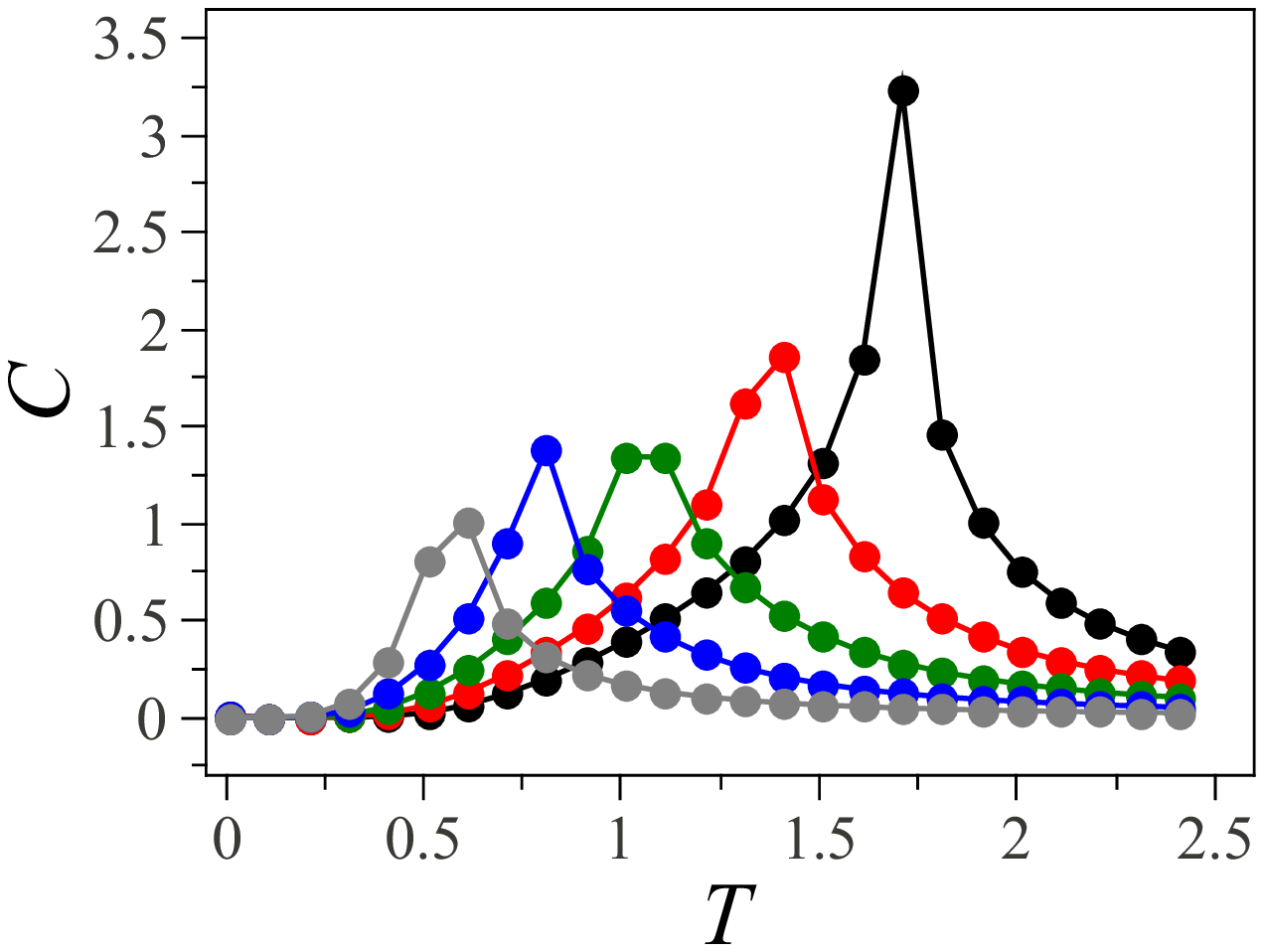}}\hfill{}\subfigure[]{\label{fig:Second-subfigure-1-1}\includegraphics[scale=0.5]{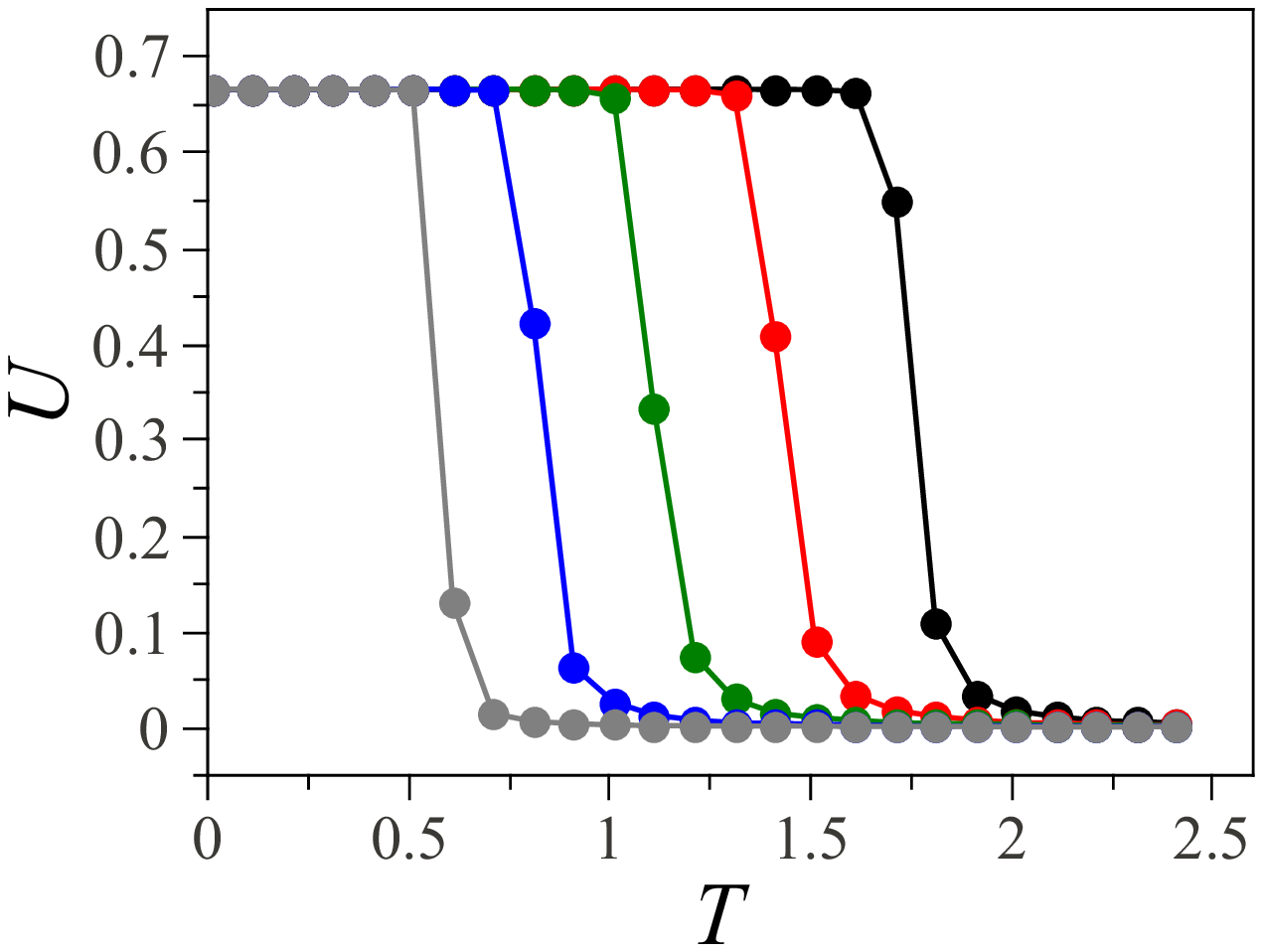}}\hfill{}

\caption{Thermodynamic quantities per site as a function of temperature $T$,
and for several values of the concentration $x$ of spin-1/2.{\small{}
(a) Magnetization $m$, (b) susceptibility $\chi$, (c) specific heat
$C$ and (d) fourth-order Binder cumulant $U$. The concentration
$x$ of spin-1/2 is $x=1.0$ (gray line), $x=0.75$ (blue line), $x=0.50$
(green line) , $x=0.25$ (red line) and $x=0$ (black line). The temperature
is measured in units of $J/k_{B}$. }}
\end{figure}

All the results of the simulations presented here were realized for
a concentration $x$ of spin-1/2 fixed at $x=$ 0, 0.25, 0.50, 0.75,
1.0. In Fig. 2, we displayed the behavior of the magnetization (Fig.
2(a)), susceptibility (Fig. 2(b)), specific heat (Fig. 2(c)) and fourth-order
Binder cumulant (Fig. 2(d)) as a function of temperature $T$, and
several values of the concentration $x$ of spin-1/2. For $x=0$,
we have the critical behavior of the Blume-Capel model with the term
of anisotropy equal zero $(D=0)$ \cite{blume,capel}. Nevertheless,
for $x=1.0$, we have the case of the pure Ising model. The critical
behavior for $x=0.5$0, i. e., half of the lattice with spin-1/2 and
another half of the lattice with spin-1 randomly distributed were
studied in the reference \cite{godoy}. We can observe that the magnetization
vanishes with the increase of temperature $T$. When the concentration
$x$ of spin-1/2 increases $(x\rightarrow1)$ the magnetization vanishes
at different values of critical temperatures $T_{c}$. The critical
temperature decreases with the increase of the concentration $x$
of spin-1/2. These can be clearly observed by the shifting in the
susceptibility and specific heat peak (see Fig. 2(b) and 2(c)).

\begin{figure}[H]
\begin{centering}
\includegraphics[clip,scale=0.5]{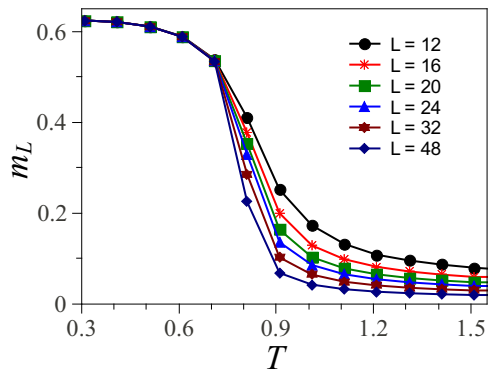} 
\par\end{centering}

\caption{{\small Magnetization $m_{L}$ as a function of temperature $T$ for
various lattice sizes $L$ as indicated in the figure. The temperature
is measured in units of $J/k_{B}$. }}
\end{figure}

The finite-size effects of the magnetization were studied for the
concentration $x$ of spin-1/2 fixed at $x=0.75$ and for various
system sizes $L$. In Fig. 3, it is shown the behavior of the magnetization
as a function of temperature $T$ for several system sizes $L$. We
can observe in this figure, that the magnetization $m_{L}$ vanishes
with the increase of temperature $T$, indicating the existence of
a phase transition.

\begin{figure}[H]
\begin{centering}
\includegraphics[scale=0.5]{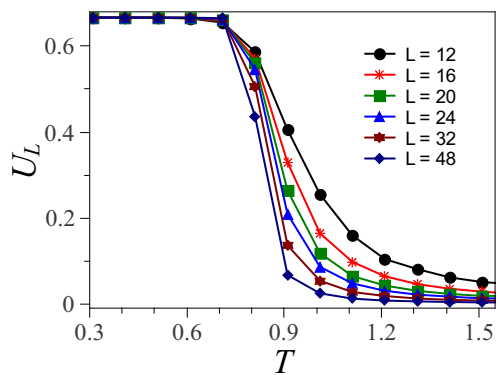} 
\par\end{centering}

\caption{{\small Fourth-order cumulants $U_{L}$ as a function of temperature
$T$ for various lattice sizes $L$ as indicated in the figure. The
temperature is measured in units of $J/k_{B}$. }}
\end{figure}

So as to study the phase transition in more details, we used the fourth-order
Binder cumulants $U_{L}$ intersection method to determine the value
of temperature at which the transition occurs. According to the theory
of finite-size scaling for continuous phase transitions, the finite-size
behavior is governed by the ratio $L/\xi$, where $\xi$ is the correlation
length. The scaling relation for the fourth-order cumulant shows that,
at critical temperature, where the correlation length is infinite,
all the curves must intercept themselves at a single point, since
$L/\xi$ is zero for all the sizes $L$. To find the critical temperature,
we displayed in Fig. 4 the cumulants $U_{L}(T)$ versus temperature
$T$, for several system sizes $L$ and for $x=0.75$. Our estimate
for the dimensionless critical temperature is $T_{c}=(0.71\pm0.01)J/k_{B}$.

The susceptibility $\chi_{L}$ as a function of temperature $T$ is
shown in Fig. 5. For finite systems $\chi_{L}$ presents a peak around
the critical temperature $T_{c}$, which grows in height with the
increase of the system size. 

\begin{figure}[H]
\begin{centering}
\includegraphics[clip,scale=0.5]{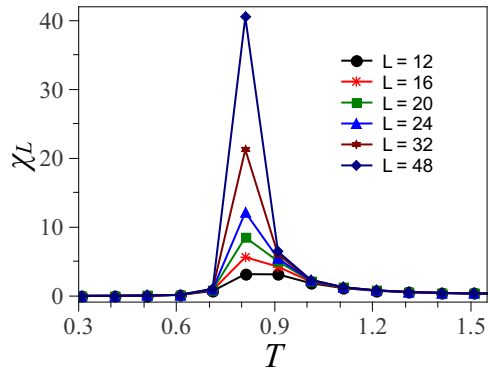} 
\par\end{centering}

\caption{{\small Susceptibility $\chi_{L}$ as a function of temperature $T$
for various lattice sizes $L$ as indicated in the figure. The temperature
is measured in units of $J/k_{B}$. }}
\end{figure}

In Fig. 6, we also present the measurements of the specific heat.
The peak observed in the curves of the specific heat exhibit a weak
system size dependence compared to the susceptibility peak. The position
of the specific heat and of the susceptibility peaks can be defined
at a pseudocritical temperature $T^{max}(L)$. The $T^{max}(L)$ approaches
$T_{c}$ when $L\rightarrow\infty$ \cite{fisher}.

\begin{figure}[H]
\begin{centering}
\includegraphics[clip,scale=0.5]{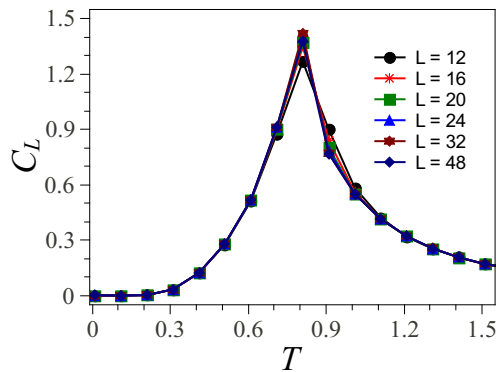} 
\par\end{centering}

\caption{{\small Specific heat $c_{L}$ as a function of temperature $T$ for
various lattice sizes $L$ as indicated in the figure. The temperature
is measured in units of $J/k_{B}$. }}
\end{figure}

We have also calculated the critical temperature $T_{c}$ for several
different values of the concentration $x$ of spin-1/2, as it can
be seen in Fig. 7. For $x=0$ (only spin-1), we found $T_{c}=(1.69\pm0.01)J/k_{B}$,
that is the critical temperature of the Blume-Capel model with $D=0.0$.
Yet, for $x=1.0$ (only spin-1/2), $T_{c}=(0.565\pm0.003)4J/k_{B}$
which is the critical temperature of the Ising model. When $x=0.50$
(half of spin-1/2 and half of spin-1) the critical temperature is
$T_{c}=(1.00\pm0.003)J/k_{B}$ , therefore, the result is not according
to the one of the reference\cite{godoy}. The general features of
the phase diagram show that the critical temperature change when concentration
varies $0<x<1.0$. 

\begin{figure}[H]
\begin{centering}
\includegraphics[clip,scale=0.5]{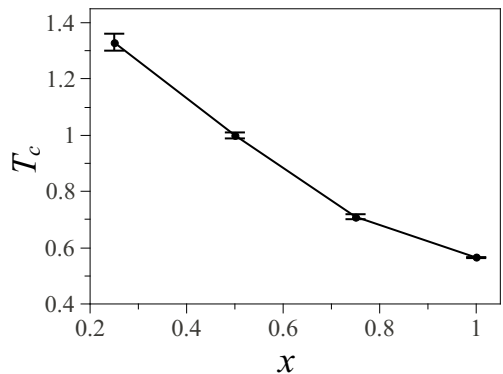} 
\par\end{centering}

\caption{{\small Critical temperatures $T_{c}$ as a function of the concentration
$x$ of spin-1/2. The critical temperatures are measured in units
of $J/k_{B}$. }}
\end{figure}

\section{Conclusions}

We have studied a disordered ferromagnetic binary Ising model on a
square lattice. The model consists of two different types of particles
with spin-1/2 and spin-1. These particles are randomly distributed
on the square lattice, and we considered only nearest-neighbor interactions.
We also calculated the critical temperature $T_{c}$ for several different
values of the concentration $x$ of spin-1/2. For $x=0$ (only spin-1),
we found the critical temperature of the Blume-Capel model with $D=0.0$.
On the other hand, for $x=1.0$ (only spin-1/2) we found the critical
temperature of the Ising model. We have showed that the phase diagram
present different critical temperatures for several values of the
concentration $x$ of spin-1/2.
\begin{acknowledgments}
This work was partially supported by the Brazilian agencies CAPES,
CNPq and FAPEMAT. \end{acknowledgments}


\begin{thebibliography}{References}
\bibitem{stinchcombe}R. B. Stinchcombe in \emph{Phase Transitions
and Critical Phenomena}, edited by C. Dombo and J. L. Lebowitz (Academic,
London, 1983), Vol. 7.

\bibitem{thorpe}M. F. Thorpe and A. R. McGurn, \emph{Phys. Rev. B}
\textbf{20}, 2142 (1979).

\bibitem{katsura}S. Katsura, \emph{Can. J. Phys.} \textbf{52}, 120
(1974).

\bibitem{kaneyoshi1}T. Kaneyoshi,\emph{ Phys. Rev. B} \textbf{34},
7866 (1986).

\bibitem{scholten}P. D. Scholten, \emph{Phys. Rev. B} \textbf{32},
345 (1985).

\bibitem{kaneyoshi2}T. Kaneyoshi, \emph{Phys. Rev. B} \textbf{33},
7688 (1986).

\bibitem{kaneyoshi3}T. Kaneyoshi, \emph{Phys. Rev. B} \textbf{39},
12134 (1989).

\bibitem{plascak}J. A. Plascak, \emph{Physica A} \textbf{198}, 655
(1993).

\bibitem{godoy}M. Godoy and W. Figueiredo, \emph{Int. J. Mod. Phys.
C}\textbf{ 20}, 47 (2009).

\bibitem{heermann}K. Binder and D. W. Heermann, \emph{Monte Carlo
Simulation in Statistical Physics. An Introduction,} 3rd edn. (Springer,
Berlin, 1997).

\bibitem{binder}K. Binder in \emph{Finite-Size Scaling and Numerical
Simulation of Statistical Systems}, edited by V. Privman (World Scientific,
Singapore, 1990) p. 173.

\bibitem{metropolis}N. Metropolis, A. Rosenbluth, M. Rosenbluth,
A. Teller and E. Teller, \emph{J. Chem. Phys.} \textbf{21}, 1087 (1953).

\bibitem{fisher}M. E. Fisher. \emph{Proceedings of the 51st Enrico
Fermi Summer School, }Varena, Italy, ed. M. S. Green, (Academic Press,
New York, 1971).

\bibitem{young}2 D.P. Belanger and A.P. Young,\emph{ J. Magn. Magn.
Mater.} \textbf{100}, 272 (1991). 

\bibitem{blume}M. Blume, \emph{Phys. Rev.} \textbf{141}, 517 (1966). 

\bibitem{capel}H. W. Capel, \emph{Physica} \textbf{32}, 966 (1966).\end{thebibliography}
\end{document}